\documentstyle[12 pt] {article}
\begin{document}
\begin{center}
{\bf Tautomeric mutation: A quantum spin modelling}
\end{center}
\begin{center}
{Ranjan Chaudhury}\\
{S.N. Bose National Centre For Basic Sciences}\\
{Block-JD, Sector-3, Salt Lake}\\
{Calcutta-700098, India}\\
 ranjan@bose.res.in; $ranjan_{021258}@yahoo.com$
\end{center}
\vspace*{2 cm}

 A quantum spin model representing tautomeric mutation is proposed for any DNA molecule. Based on this model, the quantum mechanical calculations for mutational rate and complementarity restoring repair rate in the replication processes are carried out. A possible application to a real biological system is discussed.\\
\vspace*{15 cm}

\section {Introduction}

 It is well known by now that in the cellular environment, DNA molecules are not absolutely stable and invariant. Rather each base pair in a DNA double helix has a certain probability of undergoing change. Such events occurring within a gene are known as "gene mutations"$[1]$. Moreover DNA can be regarded as a system undergoing a dynamic competition between the chemical processes leading to new mutations and the cellular repairing processes correcting these premutational events $[1,2]$.

 Here we would like to present a possible model and calculational scheme corresponding to a non-enzymatic partial repairing process leading to "complementarity restoration" for a special kind of point mutation known as "tautomeric mutation"[1,2]. The origin of this type of mutation is quantum fluctuation occurring spontaneously or at times initiated by the presence of mutagens (base analogs). In this mutational process each nucleotide (viz. A,T,C,G) can fluctuate between the two states viz. "keto" and "enol" driven by "proton tunnelling" [2]. Tautomerization is known to affect approximately $0.01$ percent of the DNA bases (nucleotides) [2]. Out of these two states the most commonly occurring keto states obey the conventional Watson-Crick base pairing rule (BPR), whereas the rare enol states can have various anomalous pairings [1,2]. During replication the presence of this enol form (if present in any of the strands) leads to anomalous base pairing and hence causes "tautomeric mutation" (TM). After the completion of more than one round of replication however, the original base pairs, involved in  normal pairing on the two strands, may change over to a new normal base pair leading to "base pair transformation" (BPT); Or else they may come back to the original unchanged base pair itself. This phenomenon is called "complementarity restoration" (CR).  

 In this communication we would like to calculate the CR rate in such a TM process. Our calculation is quantum mechanical and based on a quantum spin modelling of the DNA configurations. The model we are adopting is an analog of a 2-legged Ising spin ladder [3-6] involving two kinds of "magnetic atoms" in the presence of a "fictitious transverse field", with only inter-strand interactions. This inter-strand antiferromagnetic interaction  of the Ising type operating between the "spins" on the "like atoms" only, tries to preserve the "Watson-Crick base pairing rule (BPR)" or the "complementarity" on the two strands of the DNA. Furthermore, these two kinds of atoms themselves are also represented by another kind of "pseudo-spin" $\frac{1}{2}$ objects experiencing only an inter-strand "pseudo-ferromagnetic interaction" of Ising type, to ensure the presence of like atoms on the two strands. The intra-strand process modelled on the assumed coupling of both "spin" and "pseudo-spin" degrees of freedom to a fictitious transverse  field, tends to cause "mutational fluctuations" leading to violation of the BPR. We set up the equations for calculating the rates for both BPR violating point mutations as well as CR processes during the replication, for any arbitrary DNA sequence containing $N$ nucleotides on each strand. From this we estimate the efficiency of the "secondary repair mechanism" for a typical DNA molecule with some known parameters. We suggest a possible system and experiments to test our theoretical prediction.\\

 It may be mentionworthy that simple transverse Ising model has been used to study proton tunnelling in a double well potential configuration by many workers [3,4]. Our model and approach may therefore be looked upon as a non-trivial extension of this simple idea to the biological processes at the molecular level.\\

\section{Mathematical Formulation}

 We  model "point mutations" of tautomeric type by a quantum spin model. As mentioned in the last section, the model we employ is analogous to a transverse field  Ising model of a 2-legged ladder [3-7]. More explicitly, it is given as:-

\begin{equation}
{\mathcal{H}} = {\mathcal{H}}_{Ising}^{inter-strand rung} + {\mathcal{H}}_{Trans}^{intra-strand}
\end{equation}

 Where,\\

\begin{equation}
{{\mathcal{H}}_{Ising}} = {\sum_{i,M,N,s,s^{\prime}}}{\lambda_{i}}^{1,2} {{\sigma}^{z}}_{i,M,s;1}{{\sigma}^{z}}_{i,N,s^{\prime};2}{\delta_{MN}}{\delta_{s,-s^{\prime}}}
\end{equation}

 with the variables $M$ and $N$ both representing varieties of atoms $X$ and $Y$ and $\sigma^{z}$ taking values $s$ and $s^{\prime}$, the "spin values" corresponding to these atoms. Both $s$ and $s^{\prime}$ can assume numerical values ${\frac{1}{2}}$ and ${\frac{-1}{2}}$. The indices $1$ and $2$, occurring as superscripts and subscripts in $\lambda$ and $\sigma$ respectively, represent the two strands. The inter-strand coupling constant ${\lambda_{i}}^{1,2}$ connecting the nucleotides on the $ith$ pair of sites on the two strands, is assumed to be -ve. The nucleotides A,T,C and G can then be identified with the states $(X,\frac{1}{2})$, $(X,\frac{-1}{2})$, $(Y,\frac{1}{2})$ and $(Y,\frac{-1}{2})$ respectively. Moreover, $X$ and $Y$ are themselves represented as pseudo-spin "up state" and pseudo-spin "downstate" respectively.\\

 We have assumed in the above model that the intrinsic coupling is purely inter-strand and that the intra-strand intrinsic coupling is zero. This inter-strand coupling is antiferromagnetic for the spins  but ferromagnetic for the pseudo-spins, as is obvious from the structure of the above hamiltonian. This tries to preserve the BPR allowed nucleotide pairings on the two strands. Each strand is individually and internally "paramagnetic"(uncorrelated in both spin and pseudo-spin) but they are mutually both "spin and pseudo-spin correlated (paired)". In this model the BPR violating mutations are generated by the transverse field (assumed to be external) acting on each strand. This part of the Hamiltonian is given by:-\\
\begin{equation}
{\mathcal{H}}_{trans}^{intrastrand}={\sum_{i,l} \frac{1}{2} H^{l}_{i}[(\sigma^{+}_{i,l} + \sigma^{-}_{i,l})_{sp} X I_{psp} + I_{sp} X (\sigma^{+}_{i,l} + \sigma^{-}_{i,l})_{psp} +\frac{1}{2}(\sigma^{+}_{i,l} + \sigma^{-}_{i,l})_{sp} X (\sigma^{+}_{i,l} + \sigma^{-}_{i,l})_{psp}]}
\end{equation}
 where $sp$ stands for "spin" and $psp$ stands for "pseudo-spin". The $\sigma$'s are the Pauli matrices and $I$ is an unit matrix of $2X2$ dimension. It should be highlighted that our model differs from the conventional transverse Ising model in two major aspects:- (i) the transverse field $H^l_{i}$ couples to both spin and pseudo-spin locally and (ii) both the couplings may be present simultaneously. Moreover the site dependence of the transverse field ensures the different amplitudes for various mutational channels.\\

 Let us assume that the magnitude of $\lambda_i$ is much greater than that of $H_i$. We then attempt to treat the 2nd term in the above Hamiltonian, as a perturbation and use the ground state of the  1st term (in the Hamiltonian) as the reference state. The ground state is a configuration which obeys BPR. We confine ourselves to the case of zero temperature in the absence of any external stimuli (like UV radiation). In our scheme the order of perturbation corresponds to the order of replication of the DNA. The perturbation calculation will have to be based on the degenerate perturbation theory, as the ground state has a high degree of degeneracy, viz. $4^N$, where $N$ is the total number of nucleotides in any one strand of the DNA molecule. Let $|\psi_{i}\rangle$ and $|\psi_{j}\rangle$ be the two degenerate ground states and $|\psi^{ex}\rangle$ be the excited state. Thus for our double-stranded DNA molecule, $|\psi_{i}\rangle$ is a "parent" molecule state, $|\psi^{ex}\rangle$ represents one of the "daughter" molecule product states and $|\psi_{j}\rangle$ is the "grand daughter" molecule state. The daughter state is obtained from the parent state after the 1st round of replication causing some mutation. Again the grand daughter state is obtained from the parent state after the 2nd round of replication and which is one of the doubly mutated states degenerate with the  parent state. Thus after even number of replications (even orders of perturbation), the successive mutations can sometimes be mutually compensating to get back a new state satisfying BPR but related to the original parent state by a BPT. The other possibility is to come back to the original state itself. The odd number of replications however, always generate some BPR violating defective configurations. Besides for a real DNA molecule, even after the 1st round of replication there are normal (unmutated) configurations produced which will also continue to lead to the original parental state after the next (2nd) round of replication as well. Our model however does not include these normal replication processes. We only model the processes which involve both mutation and replication. Moreover we only allow single (point) mutations one at a time and neglect the doubly (and higher) excited states in our calculations.\\
 We would like to study here the lowest order processes viz. the second order replication for calculating the "complementarity restoration" rate (CRR) and the first order replication for calculating the tautomeric mutation rate (TMR). For these calculations, we will follow the standard procedure from quantum mechanics [8].\\ 
 
 We would need the following two quantities for our calculations:-\\

\begin{equation}
 F1(i,ex)= \langle\psi_{i}| {\mathcal{ H}} _{Trans}^{IS}|\psi^{ex}\rangle
\end{equation}
 and\\

\begin{equation}
 F2(i,j)=\sum_{ex}{\langle\psi_{i}|{\mathcal{ H}}_{Trans}^{IS}|\psi^{ex}\rangle\langle\psi^{ex}|{\mathcal{ H}}_{Trans}^{IS}|\psi_{j}\rangle}
\end{equation}

 where $|\psi_{i}\rangle$ is the "parent state", $|\psi^{ex}\rangle$ is a "daughter product state" and $|\psi_{j}\rangle$ is a "grand daughter product state", as have been discussed earlier. The symbols $i$, $j$ and $ex$ denote the indices corresponding to the two degenerate ground states viz. $|\psi_{i}\rangle$ and $|\psi_{j}\rangle$ respectively and the excited state $|\psi_{ex}\rangle$. Both $|\psi_{i}\rangle$ and $|\psi_{j}\rangle$ belong to the manifold containing $4^N$ degenerate ground states, where $N$ is the total number of nucleotides on any one strand of the DNA molecule. Again for a chosen ground state, the degree of degeneracy of the lowest excited states (containing single mutation) is $2N$, as can be seen by combining mutation with replication. Moreover for a definite $|\psi_{i}\rangle$ the grand daughter state $|\psi_{j}\rangle$ has only 2 choices for the complementarity restoration processes, corresponding to each mutated daughter state $|\psi^{ex}\rangle$. As discussed earlier, one of these two states corresponds to the original parent state and the other one a state obtained through BPT. This leads to the total number of distinct grand daughter states as $2N+1$.\\ 

 These quantities $F1$ and $F2$ will be the important ones in the calculations involving the first order and the second order processes respectively. The rate for the base pair transformation can be calculated by making use of the Fermi Golden Rule [8], as the parent state and the grand daughter state are degenerate. The calculation for BPR violating mutational rate is however more complicated, as the process does not conserve energy strictly. The deviation from the energy conservation is however $\frac{1}{N}$ times the absolute magnitude of the ground state energy. Thus for a macroscopic size of the DNA molecule, even the BPR violating single mutation conserves energy approximately. Therefore we can estimate the rate of this process by applying Fermi Golden Rule once again.  We derive an approximate expression for the rate in this case.\\
\begin{equation}
{{\tau}^{-1}}_{NBPR}^{Total}=\frac{(2\pi)^{2}\sum_{i,f}{\rho_{i}|F1(i,f=ex)|^{2}\delta(\omega_{fi})}}{h}
\end{equation}
 Here, $\rho_{i}$ is the probability of occupation of the initial (chosen) ground state with index $i$, $\omega_{fi}$ is simply $\frac{{\Delta}E}{\frac{h}{2\pi}}$, $f$ is the index of the final (lowest excited) state and ${\Delta}E$ is the energy difference between the excited state and the ground state. As has been pointed out before, for the single mutation $\frac{\Delta E}{E_{ground}}$ tends to zero. It can be seen very easily that in general $\rho_{i}$ is independent of $i$ and is equal to $4^{-N}$. Similarly, the rate for CR is given by
\begin{equation}
{{\tau}^{-1}}_{BPR}^{Total}=\frac{(2\pi)^{2}{\rho}\sum_{i,j}|F2(i,j)|^{2}\delta(\omega_{ij})}{h}
\end{equation} 
 From the above two expressions, we see that the rates for both the processes contain the matrix element of ${\mathcal H}_{Trans}^{IS}$ between a ground state and an excited state configuration. Thus it should be possible to relate the two rates and thereby estimate the repairing efficiency ($RE$). By taking the ratio of the CR rate and the TM rate, with the proper degeneracy factors, we get the expression for $RE$. Assuming an uniform external field $H$ for simplicity, the expression for $RE$ becomes equal to $2H^{2}$.\\

\section{Calculations and Results}

  As an example we now estimate the magnitude of the fictitious transverse field $H$, the simulator of mutation introduced in our model, from the available data for spontaneous point mutation corresponding to a typical DNA molecule. It is found that in the absence of mutagenic treatment the effective spontaneous mutation rate (after primary repair processes or "proof reading") is about $2X10^{-11}$ per second per nucleotide for a well studied organism (a fungus) named "Neurospora"[1]. Since in this case the parental DNA configuration and the mutation channels are all fixed, assuming that all the mutations referred above are of tautomeric type having taken place in the 1st round of replication itself, we get the following equation involving the rate:- 
\begin{equation}
 H^{2}n = NX2X10^{-11}
\end{equation}
 where $n$ is the number of mutation channels (mutated daughter states)  allowed after the 1st round of replication, for this species. As mentioned before, $n$ is proportional to $N$ and in fact is equal to $2N$. Thus the magnitude of $H$ becomes approximately $3X10^{-5}$ in the suitable unit and is essentially independent of $N$. This leads to the magnitude of the secondary repairing efficiency of the order of $10^{-11}$. To be more precise, according to our calculation about 2 out of every $10^{11}$ point mutations  are compensated (complementarity is restored) through the secondary repairing processes arising from quantum fluctuation itself, for Neurospora. Thus this number will change with different choices of the species only if the rate of mutation changes [9].\\
 The results will be very different however, if we start from an arbitrary DNA configuration corresponding to a given number of nucleotides $N$ per each strand. In that case, the rate equation for mutation becomes,
\begin{equation}
2H^{2}(4^{-N}) = r
\end{equation}
 where $r$ is the rate of mutation per nucleotide. In this case, the full degeneracy of the parent state as well as that of the daughter state have been taken into account. This leads to a very large magnitude of $H$ even for a moderate value of $N$ and reasonable value of $r$. Thus this violates the condition that $H$ will have to be much less than $\lambda$ for the perturbation treatment to be valid. Therefore our formalism and analysis are appropriate for DNA sequences corresponding to any definite species only. Moreover according to our prediction,  the secondary repairing efficiency turns out to be of the same order of magnitude  as the single mutation (point mutation) rate itself. This interesting result will have to be verified from the observational data after filtering out the tautomeric mutational component and taking into consideration the rate of production of the total $4N$ grand daughter configuration states generated after the 2nd round of replication from the singly mutated daughter states. This comparison will also throw some light on the validity of our modelling as well as the calculational scheme for studying the process of complementarity restoration. It may be remarked that amongst these $4N$ states, the one half of them would be the original parent state and the other half the base pair transformed states. Therefore we have totally $(2N+1)$ distinct grand daughter states which are the products of the complementarity restoration process.
   
\section{Conclusions}

 We have attempted a modelling of the replication accompanied by tautomeric mutation in DNA, by constructing a  quantum spin-pseudospin based model analogous to some of the well known models used in statistical mechanics and magnetism [3-7,10-12]. The idea has been to incorporate quantum fluctuations to the existing ordered structure in a double-stranded DNA and calculate the rates of various processes. In particular, we considered both the disordering (mutation) as well as the order restoring CR events and calculated the efficiency of the restoring processes. It should also be highlighted that our result regarding the RE, is independent of the numerical magnitudes of the intrinsic coupling constants appearing in our model [9].\\

 We have tried to apply our formalism and calculations to an experimental biological system. However, due to non-availability of sufficient data the comparison of observational results with our prediction remains incomplete at present. To summarize, in our work we have chalked out a scheme for a possible novel repair mechanism for the DNA mutations, which is based entirely on quantum fluctuations. This is distinct from the conventional processes viz. primary repair system and the post-replication repair mechanism which are enzyme initiated [1]. It should be made clear however, that our scheme is applicable only to the case of tautomeric mutations.\\

 The possible future experiments should consist of detailed rate measurement of the processes of mutation as well as complementarity restoration arising purely out of tautomerization. For this the conventional methods like that of Luria and Delbruck [1,13] may have to be modified. This should enable one to critically examine our modelling and prediction and compare our theoretical results more precisely with the relevant observations. On the theoretical side too, there are lot of scopes for improvement and making the model more realistic by taking into account the intrinsic intra-strand couplings as well as the site and strand dependence of the transverse field. Last but not the least, this model suggested by us, although partially borrowed from condensed matter physics, has a lot of potentiality to describe the microscopic processes occurring within a DNA. This attempt is further inspired by our previous success with the analysis of the scaling behaviour of DNA involving the nucleotide distribution, by the methodologies of the statistical physics [14].\\ 

\section{References}
 1. Griffiths A.J.F., Gelbart W.M., Lewontin R.C. and Miller J.H., in 'Modern Genetic Analysis: Integrating Genes and Genomes', edited by Noe J. et al (W.H. Freeman and Company, New York) 2002, chap.10, pp. 313-348; chap.2, pp. 31.\\
 2. McFadden J., in 'Quantum Evolution: Life in the Multiverse', (Flamingo publishers, London) 2000, Chap.3, pp. 49-66.\\
 3. Henkel M., Harris A.B. and Cieplak M., Phys. Rev. B, 52 (1995) 4371.\\
 4. Barreto F.C.Sa, Braz. J. Phys., 30 (2000).\\
 5. Prakash S., Havlin S., Schwartz H. and Stanley H.E., Phys. Rev. A, 46 (1992) R1724.\\
 6. Scalapino D.J., Zhang S.C. and Hanke W., Phys. Rev. B, 58 (1998) 443.\\
 7. Turban L., J. Phys. A, 15 (1982) 1733.\\
 8. Schiff L.I., in 'Quantum Mechanics', (McGraw-Hill Book Company, Singapore) 1968, Chap.8, pp. 283-285.\\
 9. Yakushevich Ludmila V., in 'Nonlinear Physics of DNA', (Wiley-Vch Verlag GmbH and Co. KGaA, Weinheim) 2004, Chap.1, pp.4-7; Chap.3 pp.47.\\
 10. Song J., Gu S. and Lin H., eprint arxiv: quant-ph/0606207, preprint, 2006.\\
 11. Laad M.S. and  Lal S., arxiv: cond-mat/0510016,v2, preprint, 2005 (To appear in Europhys. Lett.).\\
 12. Chaudhury Ranjan, J. Mag. and Magn. Mat., 307/1 (2006) 99.\\
 13. Coulondre C., Miller J.H., Farabaugh P.J. and Gibert W., Nature, 274 (1978) 775.\\
 14. Som A., Sahoo S., Mukhopadhyay I., Chakrabarti J. and Chaudhury R., Europhys. Lett., 62 (2003) 277.\\

\end{document}